\documentclass[pra,aps,twocolumn,showpacs,amsmath, amssymb]{revtex4}

\usepackage{graphicx}
\usepackage{dcolumn}
\usepackage{bm}
\usepackage{color}

\begin{document}

\title{Vortex splitting and phase separating instabilities of coreless vortices in $F = 1$ spinor Bose-Einstein condensates}

\author{M. Takahashi$^1$, V. Pietil\"a$^{2, 4}$, M. M\"ott\"onen$^{2, 3, 4}$, T. Mizushima$^1$, and K. Machida$^1$}

\affiliation{$^1$Department of Physics, Okayama University, Okayama 700-8530, Japan}
\affiliation{$^2$Department of Applied Physics/COMP, Helsinki University of Technology, P. O. Box 5100, FI-02015 TKK, Finland}
 \affiliation{$^3$Low Temperature Laboratory, Helsinki University of Technology, P. O. Box 3500, FI-02015 TKK, Finland}
 \affiliation{$^4$Australian Research Council, Centre of Excellence for Quantum Computer Technology, The University of New South Wales, Sydney 2052, Australia}

\date{\today}

\begin{abstract}
The low lying excitations of coreless vortex states in $F \! = \! 1$
spinor Bose-Einstein condensates (BECs) are theoretically investigated using 
the Gross-Pitaevskii and Bogoliubov-de Gennes equations. 
The spectra of the elementary excitations are
calculated for different spin-spin interaction parameters and
ratios of the number of particles in each sublevel.
There exist dynamical instabilities of the vortex state
which are suppressed by ferromagnetic interactions,
and conversely, enhanced by antiferromagnetic interactions.
In both of the spin-spin interaction regimes,
we find vortex splitting instabilities in analogy with scalar BECs.
In addition, a phase separating instability is found
in the antiferromagnetic regime.
\end{abstract}
\maketitle

\section{Introduction}
The realization of atomic Bose-Einstein condensates
(BECs)~\cite{mh_anderson, bradley, davis, pethick} constituted the
beginning of a new era in atomic physics. Compared with the
traditional solid state systems, BECs in ultracold atomic gases
have several appealing features, such as tunable interaction
strengths, various trap potentials, and direct observation of the
particle density. These properties offer a unique venue for many
different types studies such as the stability of the BECs in a
trap potential, topological defects, multi-component BECs, BECs in
optical lattices, and low dimensional Bose gases~\cite{varenna,
les_houches, bloch}.

In this work, we focus on vortex states in spinor
BECs~\cite{ohmi,ho} which have been realized experimentally in
$^{23}$Na and $^{87}$Rb with the hyperfine spin states $F \! = \!
1$ and $F \! = \! 2$ \cite{stenger, gorlitz, barrett,
schmaljohann, chang, kuwamoto}. In these experiments, the atoms
are confined optically and the condensate exhibits a genuine spin
degree of freedom. Due to the $(2 F \! + \! 1)$ different
hyperfine spin sublevels, the spinor BECs can have several
topologically different stationary states, including vortex states.
The quantized vortex is defined as a phase singularity of the condensate wave function \cite{pethick}.
The phase of the wave function winds by $2 \pi n$ about the vortex
core, where the integer $n$ is referred to as the vortex quantum
number. For scalar BECs, the particle density vanishes at the
vortex core due to diverging superfluid velocity. Due to the
internal states in the spinor BECs, several types of
vortices and other topological defects can be found. Studies
related to these topological defects have been carried out
experimentally in Refs. \cite{leanhardt, kevin}.
Theoretical studies of vortices and other topological defects in
$F \! = \! 1$ spinor BECs were initiated by Ohmi and Machida
\cite{ohmi} and Ho \cite{ho}. Systematic investigations on
vortices were followed by Yip \cite{yip} who considered both
axisymmetric and nonaxisymmetric vortices, and by Isoshima {\it et al.}
\cite{isoshima2, isoshima4, nakahara, isoshima5},
who considered only axisymmetric vortices and their excitation
spectra. Studies of different types of related topological defects
have also been carried out in the literature: Leonhardt and Volovik
\cite{leonhardt}, studied a defect referred to as Alice which is
also known as the half quantum vortex. Stoof \cite{stoof} and
Marzlin {\it et al.} \cite{marzlin} studied so-called skyrmions.
Mizushima {\it et al.} \cite{mizushima1, mizushima2, mizushima4}
and Pietil{\" a} {\it et al.} \cite{ville} studied coreless
vortices, also known as Mermin-Ho \cite{mermin} or
Anderson-Toulouse vortices \cite{pw_anderson}. Furthermore, other
studies of the exotic properties of $F \! = \! 1$ spinor BECs have
carried out in Refs.~\cite{robins, zhou, kasamatsu}. For $F>1$, also many
theoretical studies have been reported
\cite{ciobanu, koashi, ueda, mottonen2, makela, semenoff, martikainen, pogosov}.

In this work, we consider the coreless vortex states in spinor BECs with hyperfine spin $F \! = \! 1$.
In the $z$-quantized basis, the condensate order parameter is denoted by $\phi_i$ where $i \! = \! 1, 0, -1$.
For the coreless vortex state, the core of the vortex is filled by one of the components of the order parameter $\phi_i$.
Thus the coreless vortex is fundamentally different from the
vortex in scalar BECs. Typically, the coreless vortex state can be
defined by a combination of winding numbers $\langle w_1, w_0,
w_{-1} \rangle \! = \! \langle 0, 1, 2 \rangle$ \cite{mizushima1,
ville}. However, by changing the magnetization per particle $M$,
analogous vortex states to the ones in a scalar BEC can be
realized in the limit $M \! = \! -1$, since in this case the state
$\phi_{-1}$ is fully populated. In this limit, the coreless vortex
state of the condensate corresponds to a doubly quantized vortex
in a scalar condensate which is known to be dynamically unstable
\cite{pu, mottonen, shin, huhtamaki1, kawaguchi, huhtamaki2, lundh,
isoshima11}.
The dynamical instability is characterized by the appearance of the complex-frequency eigenmodes in the excitation spectrum (see Sec.~\ref{sec:formulation}).
The existence of excitations with negative but real energy is referred to as energetic instability or local instability and it implies that there is a stationary state with smaller energy to which the system tends to decay in the presence of dissipation.
On the other hand, the $M \! = \! 1$ limit is a vortex-free state of a scalar BEC, which is the ground state in nonrotating systems.
It is thus expected that the nature of the instability of the coreless vortex state changes as a function of $M$ between these two extreme limits.

Let us discuss the differences between the present and previous
studies. The condensate phase diagram in a plane of $M$ and
external rotation $\Omega$ has been partially studied in Refs.~\cite{isoshima5, mizushima1, mizushima4}. Mizushima {\it et al.}
\cite{mizushima4} focused on the ground state properties in the range $0 \!
\le \! M \! \le \! 1$, and Isoshima {\it et al.} \cite{isoshima5}
studied axisymmetric vortex states with winding numbers $w_i \! <
\! 2$ in the range $-1 \! \le \! M \! \le \! 1$. However, these studies are
not focused on the dynamical instability. The dynamical
instability of the coreless vortex in a Ioffe-Pritchard magnetic
field has been studied by Pietil{\" a} {\it et al.} \cite{ville},
but only the ferromagnetic case was considered.

In this paper, we focus on the existence and characteristics of
the dynamical instabilities in multicomponent systems.
The coreless vortex state is an
advantageous choice for these studies since each limit of the
magnetization corresponds either to a dynamically unstable or
stable state of a scalar BEC. We clarify how the dynamical
instabilities of the coreless vortex state change as a function of
magnetization $M$ in both ferromagnetic and antiferromagnetic
interaction regimes.
It is topical to study the
antiferromagnetic regime since the coreless vortex state has been
realized in $^{23}$Na atoms with $F \! = \! 1$ using the
topological phase imprinting method \cite{leanhardt} according to the theoretical proposal \cite{isoshima2, nakahara}.
On the other
hand, $^{87}$Rb atoms in $F \! = \! 1$ hyperfine spin state
constitutes a ferromagnetic BEC.
We demonstrate different aspects of dynamical instabilities in
these two interaction regimes. The dynamical instability is
suppressed by the ferromagnetic interactions, whereas it is
enhanced by the antiferromagnetic interactions. In the latter
case, there are two kinds of dynamical instabilities: the vortex
splitting and phase separating instabilities. In addition, we
discuss the physical mechanisms behind these results.

This paper is organized as follows.
In Sec.~\ref{sec:formulation}, we introduce a theoretical description and details of the studied system.
In Sec.~\ref{sec:result}, we illustrate the condensate order
parameter as a function of $M$ in different spin-spin interaction
regimes and show a typical excitation spectrum including complex
eigenvalues.
Then we present our main results on the dynamical instabilities arising
for different spin-spin interactions.
In Sec.~\ref{sec:conclusion}, we conclude our study.
In the Appendix, we provide a proof of the existence of the
so-called Kohn modes in the mean-field picture of spinor BECs.

\section{System and Formulation}
\label{sec:formulation}

We begin with the second quantized Hamiltonian for an $F \! = \!
1$ spinor BEC \cite{pethick} in the absence of a magnetic field,
\begin{eqnarray}
{\hat H}
&=& \int d{\bm r} \Biggl [ \sum_i {\hat \Psi}_i^\dagger H_i^0 {\hat \Psi}_i
+ \frac{g_n}{2} \sum_{i, j} {\hat \Psi}_i^\dagger {\hat \Psi}_j^\dagger
{\hat \Psi}_j {\hat \Psi}_i \nonumber \\
&& \mbox{} + \frac{g_s}{2} \sum_{i, j, k, l} \sum_{\alpha}
{\hat \Psi}_i^\dagger {\hat \Psi}_j^\dagger \left ( F_\alpha \right )_{i, l}
\left ( F_\alpha \right )_{j, k} {\hat \Psi}_k {\hat \Psi}_l \Biggr ],
\end{eqnarray}
where
\begin{eqnarray}
H_i^0
&=& - \frac{\hbar^2}{2 m} \nabla^2 + V_{\rm trap} ({\bm r}) - {\bf \Omega} \cdot
\left ( - i \hbar {\bm r} \times \nabla \right ) - \mu_i,
\end{eqnarray}
and $\hat{\Psi}_i$ is the bosonic field operator in the $i$th spin
sublevel and $m$ is the mass of the atoms. Here $\left ( F_\alpha
\right )_{i, j}$ is the ($i$, $j$) component of the spin matrix
$F_\alpha$ ($\alpha \! = \! x, y, z$) for hyperfine spin $F \! =
\! 1$ system. The chemical potential is defined as $\mu_j \! = \!
\mu + j \delta \mu$ in our calculation. The
subscripts $\{ i, j, k,
l \}$ take values of the spin sublevels $1$, $0$, and $-1$. The
strength of the density-density and spin-spin interactions are
denoted by the coupling constants $g_n \! = \! 4 \pi \hbar^2 ( a_0
\! + \! 2 a_2 ) / 3 m$, and $g_s \! = \! 4 \pi \hbar^2 (a_2 \! -
\! a_0 ) / 3 m$, respectively. Here $a_0$ and $a_2$ are the
$s$-wave scattering lengths between atoms with total spin $0$ and
$2$, respectively.
In our calculation,
the axisymmetric trap potential is
$V_{\rm trap} ({\bm r})\! = \! \frac{1}{2} m \omega^2 r^2$ with $r \! = \! \sqrt{x^2 \! + \! y^2}$
and the external rotation is taken along the $z$ axis ${\bm \Omega} \! = \! ( 0, 0, \Omega )$. We consider a uniform system along the $z$ direction.

Following the standard procedure \cite{ohmi, ho}, we write the time-dependent Gross-Pitaevskii (TDGP) equation as
\begin{eqnarray}
i \hbar \frac{\partial {\tilde \psi}_i ({\bm r}, t)}{\partial t}
&=& \left [ H_i^0 + g_n \sum_j | {\tilde \psi}_j ({\bm r}, t) |^2 \right ] {\tilde \psi}_i ({\bm r}, t)
\nonumber \\
&& \makebox[-25mm]{} + g_s \sum_{j, k, l} \sum_\alpha
\left ( F_\alpha \right )_{j, l} \left ( F_\alpha \right )_{i, k}
{\tilde \psi}_j^\ast ({\bm r}, t) {\tilde \psi}_k ({\bm r}, t) {\tilde \psi}_l ({\bm r}, t).
\label{eq:gp}
\end{eqnarray}
Here, the field operator $\hat{\Psi}_i$ has been replaced by its
expectation value $\tilde{\psi_i} ({\bm r}, t) \! = \! \langle
\hat{\Psi}_i ({\bm r}, t) \rangle$.
In our simulations, we find the stationary state $\phi_i ({\bm
r})$ using imaginary time propagation.

In an axisymmetric configuration, the wave function can be decomposed into the amplitude and phase factor as
\begin{eqnarray}
\phi_i ({\bm r})
= \phi'_i (r) \gamma_i (\theta) = \phi'_i (r) \exp[ i ( \alpha_i + w_i \theta )].
\label{eq:symeq}
\end{eqnarray}
Following the arguments of Isoshima {\it et al.} \cite{isoshima5}, stationary states obey conditions
\begin{eqnarray}
2 \alpha_0 = \alpha_1 + \alpha_{-1} + n \pi,
\label{eq:phase}
\end{eqnarray}
\begin{eqnarray}
2 w_0 = w_1 + w_{-1},
\label{eq:winding}
\end{eqnarray}
with $n \! \in \! {\mathbb Z}$.
In Eq.~(\ref{eq:phase}), we choose $n \! = \! 0$, $\alpha_0 \! = \! \alpha_{\pm 1} \! = \! 0$ for the ferromagnetic interaction,
and $n \! = \! 1$, $\alpha_0 \! = \! \pi/2$, $\alpha_{\pm 1} \! = \! 0$ for the antiferromagnetic case, without loss of generality.
Hence, we can choose $\{ \phi'_i \}$ to be real positive valued functions in the following discussion. These choices have essentially no effect in the discussion below.
Note that the coreless vortex states are defined as $\langle w_1, w_0, w_{-1} \rangle \! = \! \langle 0, 1, 2 \rangle$, which satisfies Eq.~(\ref{eq:winding}).

Let us consider small fluctuations in the vicinity of the stationary state $\phi$:
\begin{eqnarray}
\psi_i ({\bm r}, t) &\!=\!& \phi_i ({\bm r}) + \lambda \left (
u_{{\bf q}, i} ({\bm r}) e^{i \frac{E_{\bf q}}{\hbar} t} - v_{{\bf
q}, i}^\ast ({\bm r}) e^{- i \frac{E_{\bf q}^*}{\hbar} t} \right
). \label{eq:linearres}
\end{eqnarray}
By linearizing Eq.~(\ref{eq:gp}) with respect to $\lambda$,
we obtain the Bogoliubov-de Gennes (BdG) equation,
\begin{eqnarray}
\hat{T} {\bf w}_{\bf q}
&=& E_{\bf q} {\bf w}_{\bf q}
\label{eq:bdg}
\end{eqnarray}
where the BdG operator $\hat{T}$ is composed of $3\times 3$ complex matrices 
${\underline P}$ and ${\underline Q}$,
\begin{eqnarray}
{\hat T}
&\equiv& \left [ \begin{array}{cc}
{\underline P} & - {\underline Q} \\
{\underline Q}^\ast & - {\underline P}^\ast
\end{array} \right ].
\label{eq:bdgmatrix}
\end{eqnarray}
The matrix elements of ${\underline P}$ and ${\underline Q}$ are given 
by~\cite{isoshima4}
\begin{eqnarray}
P_{i, j}
&=& H_i^0 \delta_{i, j} + g_n \left ( \phi_i \phi_j^\ast
 +  \sum_k | \phi_k |^2 \delta_{i, j} \right ) \nonumber \\
&& \makebox[-13mm]{} \! + \! g_s \sum_{k, l} \sum_\alpha
\left [ \left ( F_\alpha \right )_{k, j} \left ( F_\alpha \right )_{i, l}
\! + \! \left ( F_\alpha \right )_{k, l} \left ( F_\alpha \right )_{i, j} \right ]
\phi_k^\ast \phi_l,
\end{eqnarray}
\begin{eqnarray}
Q_{i, j}
&=& g_n \phi_i \phi_j + g_s \sum_{k, l} \sum_\alpha
\left ( F_\alpha \right )_{j, k} \left ( F_\alpha \right )_{i, l}
\phi_k \phi_l.
\end{eqnarray}
In Eq.~(\ref{eq:bdg}), the eigenfunction is denoted by 
\begin{eqnarray}
{\bf w}_{\bf q}
= \left [ \begin{array}{c}
{\bm u}_{\bf q} \\
{\bm v}_{\bf q} \end{array} \right ],
~
{\bm u}_{\bf q} = \left [ \begin{array}{c}
u_{{\bf q}, 1}\\
u_{{\bf q}, 0}\\
u_{{\bf q}, {-1}} \end{array} \right ],
~
{\bm v}_{\bf q} = \left [ \begin{array}{c}
v_{{\bf q}, 1}\\
v_{{\bf q}, 0}\\
v_{{\bf q}, {-1}} \end{array} \right ].
\end{eqnarray}
Since the BdG matrix~(\ref{eq:bdgmatrix}) is generally non-Hermitian,
the eigenvalues $E_{\bf q}$ can be complex.

The BdG matrix has two symmetries
\begin{eqnarray}
\hat{T}^\ast
&=& - \hat{\tau}_1 \hat{T} \hat{\tau}_1,
\label{eq:sym1}
\end{eqnarray}
and
\begin{eqnarray}
\hat{T}^\dagger
&=& \hat{\tau}_3 \hat{T} \hat{\tau}_3,
\label{eq:sym2}
\end{eqnarray}
where we have introduced the first and third Pauli matrices as
$\hat{\tau}_1 \! \equiv \! \left [ \begin{array}{cc}
\underline{0} & \underline{\tau_0} \\
\underline{\tau_0} & \underline{0} \end{array} \right ]$
and $\hat{\tau}_3 \! \equiv \! \left [ \begin{array}{cc}
\underline{\tau_0} & \underline{0} \\
\underline{0} & - \underline{\tau_0} \end{array} \right ]$, where
$\underline{\tau_0} \! \equiv \! {\rm diag} (1, 1, 1)$ and 
$\underline{0}$ is a $3\times 3$ matrix of zeros.
The first symmetry in Eq.~(\ref{eq:sym1}) 
implies the existence of  
two symmetric eigenmodes
\begin{eqnarray}
\left ( E_{\bf q}, {\bf w}_{\bf q} \right )
&\Longleftrightarrow&
\left ( - E_{\bf q}^\ast, \hat{\tau}_1 {\bf w}_{\bf q}^\ast \right ).
\end{eqnarray}
These modes have opposite angular momenta under the axial symmetry.

Using the second symmetry in Eq.~(\ref{eq:sym2}),
we obtain
\begin{eqnarray}
\left ( E_{\bf q}^\ast - E_{\bf q'} \right ) \int d{\bm r}
{\bf w}_{\bf q}^\dagger ({\bm r}) \hat{\tau}_3 {\bf w}_{\bf q'} ({\bm r})
&=& 0.
\label{eq:orthonormalization}
\end{eqnarray}
For a real eigenvalue $E_{\bf q}^\ast \! = \! E_{\bf q}$,
Eq.~(\ref{eq:orthonormalization}) implies that for ${\bf q} \neq {\bf q'}$ the 
two modes are orthogonal. Thus we use normalization 
\begin{eqnarray}
\int d{\bm r} {\bf w}_{\bf q}^\dagger ({\bm r})
\hat{\tau}_3 {\bf w}_{\bf q'} ({\bm r})
&=& \delta_{{\bf q}, {\bf q'}}.
\label{eq:orthonormalizationreal}
\end{eqnarray}
for quasiparticle amplitudes corresponding to real eigenvalues of the BdG 
equation.
Two modes provided by the symmetry in Eq.~(\ref{eq:sym1}) 
give identical contribution to the energy of the quasiparticles and thus 
only the mode with positive norm in Eq.~(\ref{eq:orthonormalizationreal}) is 
chosen as a physically meaningful mode.

In the case of the complex eigenvalue $E_{\bf q}^\ast \! \ne \! E_{\bf q}$,
Eq.~(\ref{eq:orthonormalization}) gives $\int d{\bm r} {\bf w}_{\bf q}^\dagger \hat{\tau}_3 {\bf w}_{\bf q} \! = \! 0$ and 
we take the following normalization condition~\cite{mine}
\begin{eqnarray}
\int d{\bm r} {\bf w}_{\bf q}^\dagger ({\bm r}) \hat{\tau}_3 \bar{{\bf w}}_{\bf q'} ({\bm r})
&=& \delta_{{\bf q}, {\bf q'}}.
\label{eq:orthonormalizationcomplex}
\end{eqnarray}
We use a pair of eigenmodes ($E_{\bf q}$, ${\bf w}_{\bf q}$)
and ($\bar{E}_{\bf q}$, $\bar{{\bf w}}_{\bf q}$), 
for which 
the eigenvalues satisfy
the condition $E_{\bf q}^\ast \! = \! \bar{E}_{\bf q}$.
We can find such an eigenmode
as follows.
By introducing a unitary matrix
$\hat{U} \! \equiv \! {\rm diag} [ \underline{A}, \underline{A}^\ast ]$,
where $\underline{A} \! \equiv \! e^{i \alpha_1} {\rm diag} [ 1, e^{i \alpha}, e^{2 i \alpha} ]$,
and $\alpha \! \equiv \! \alpha_0 - \alpha_1$,
which renders the BdG matrix~(\ref{eq:bdgmatrix}) real,
Eq.~(\ref{eq:bdg}) can be written in the form
\begin{eqnarray}
\hat{T}' {\bf x}_{\bf q}
&=& E_{\bf q} {\bf x}_{\bf q},
\label{eq:modbdgr}
\end{eqnarray}
where $\hat{T}' \! = \! \hat{U}^\dagger \hat{T} \hat{U}$ is a real matrix. 
The complex conjugate of the equation assumes the form
\begin{eqnarray}
\hat{T}' {\bf x}_{\bf q}^\ast
&=& E_{\bf q}^\ast {\bf x}_{\bf q}^\ast,
\label{eq:modbdgc}
\end{eqnarray}
For a real eigenvalue $E_{\bf q}^\ast \! = \! E_{\bf q}$,
the eigenfunction can be taken to be real 
${\bf x}_{\bf q}^\ast \! = \! {\bf x}_{\bf q}$.
Hence Eqs.~(\ref{eq:modbdgr}) and (\ref{eq:modbdgc}) are identical.
For a complex eigenvalue $E_{\bf q}^\ast \! \ne \! E_{\bf q}$,
eigenfunction ${\bf x}_{\bf q}$ is complex, i.e.,
${\bf x}_{\bf q}^\ast \! \ne \! {\bf x}_{\bf q}$.
The eigenfunctions in Eq.~(\ref{eq:modbdgr}) and (\ref{eq:modbdgc}) are
${\bf x}_{\bf q} \! = \! \hat{U}^\dagger {\bf w}_{\bf q}$ for $E_{\bf q}$ and
${\bf x}_{\bf q}^\ast \! = \! \hat{U} {\bf w}_{\bf q}^\ast$ for $E_{\bf q}^\ast$,
where we used $U^\ast \! = \! U^\dagger$.
Here $(E_{\bf q},{\bf w}_{\bf q})$ is a solution of the eigenvalue 
equation~(\ref{eq:bdg}).
For a complex $E_{\bf q}$, the eigenstates in Eq.~(\ref{eq:bdg}) appear as a 
pairs $(E_{\bf q},{\bf w}_{\bf q})$ and $(E_{\bf q}^\ast,\bar{\bf w}_{\bf q})$, 
where $\bar{\bf w}_{\bf q} \! = \! \hat{U}^2 {\bf w}_{\bf q}^\ast$.
The additional factor $\hat{U}^2$ in $\bar{\bf w}_{\bf q}$
changes the relative phase between spin components.

By choosing the normalization condition Eq.~(\ref{eq:orthonormalizationcomplex})
for complex eigenmodes,
one can construct a complete set with the complex-frequency modes \cite{mine}.
The normalization condition in Eq.~(\ref{eq:orthonormalizationcomplex}) leaves 
the relative amplitude between ${\bf w}_{\bf q}$ and $\bar{\bf w}_{\bf q}$ as 
well as their phase undetermined.
In our study, we take equal amplitudes for ${\bf w}_{\bf q}$ and 
$\bar{\bf w}_{\bf q}$, that is,  
$|u_{{\bf q}, i} ({\bm r}) | \! = \! |\bar{u}_{{\bf q}, i} ({\bm r}) |$
and $|v_{{\bf q}, i} ({\bm r}) | \! = \! |\bar{v}_{{\bf q}, i} ({\bm r}) |$.
The physical interpretation of the quasiparticle amplitudes 
${\bf w}_{\bf q}$ and $\bar{{\bf w}}_{\bf q}$
corresponding to a complex eigenvalue is still an open question \cite{sunaga}.

\begin{figure}[h]
\includegraphics[width=\linewidth]{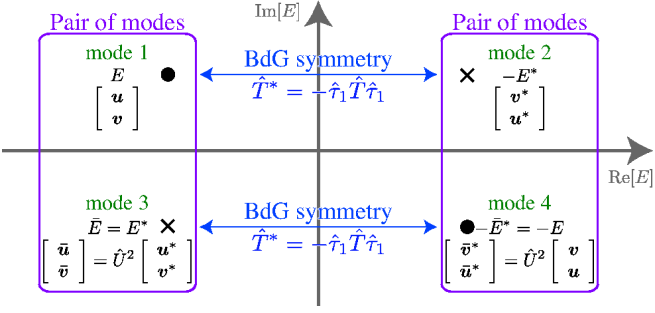}
\caption{(color online) A set of complex-frequency modes in the complex plane.
Four complex-frequency modes exist together.
}
\label{fig:complexeigenvalue}
\end{figure}

The summary of the complex-frequency modes are shown in Fig.~\ref{fig:complexeigenvalue},
where we omit quantum indices in the figure.
Modes 1 and 2 as well as modes 3 and 4 are linked by the symmetry in 
Eq.~(\ref{eq:sym1}).
Modes 1 and 3, and modes 2 and 4 in Fig.~\ref{fig:complexeigenvalue} are  
used to construct the normalization condition in 
Eq.~(\ref{eq:orthonormalizationcomplex})
for a complex eigenvalue $E_{\bf q}^\ast \! \ne \! E_{\bf q}$.
For the complex-frequency eigenmodes, the two modes which satisfy
the conservation of the energy and angular momentum are in resonance
with each other (See Sec.~\ref{subsec:excitation} for details).
For an axially symmetric system, all the eigenmodes of Eq.~(\ref{eq:bdg}) can be classified with the quantum number $q_\theta \! \in \! {\mathbb Z}$ denoting the angular momentum with respect to the condensate. The eigenfunction is thus of the form
\begin{eqnarray}
u_{{\bf q}, i} ({\bm r})
&=& u_{{\bf q}, i} (r) \exp[ i (q_\theta + w_i) \theta],
\label{eq:u}
\end{eqnarray}
\begin{eqnarray}
v_{{\bf q}, i} ({\bm r})
&=& v_{{\bf q}, i} (r) \exp[ i (q_\theta - w_i) \theta ].
\label{eq:v}
\end{eqnarray}
We solve the BdG equation using the decomposition of Eqs.~(\ref{eq:u}) and (\ref{eq:v}) to obtain the spectrum of the low-lying excitations.

From this point on, we use dimensionless quantities. The energy is normalized by the trap energy $\hbar \omega$, and the length is normalized by $d \! \equiv \! \sqrt{\hbar / m \omega}$.
In our study, we choose the density-density coupling constant to $g_n' \! \equiv \! g_n / (\hbar \omega d^3) \! = \! 0.113$, and spin-spin coupling constant to $g_s' \! \equiv \! g_s / (\hbar \omega d^3) \! = \! \pm 0.001$, $\pm 0.01$.
The negative values of $g_s$ correspond to the ferromagnetic case
and the positive ones to the antiferromagnetic case.
The values of the coupling constants $g_n'$ and $g_s'$ in the physical system \cite{klausen, crubellier} can be varied by tuning the trap frequency.
In addition, $g_n$ can be changed by using Feshbach resonances,
and hence the ratio of $g_n$ and $g_s$ is also adjustable.
We note that a drawback in utilizing the standard dc Feshbach resonance is
that it tends to fix the magnetization of the cloud because of a
required strong magnetic field.
We assume an infinitely long axisymmetric system along the $z$ axis, which renders the numerical problem two dimensional.
Alternatively, our results apply to pancake-shaped condensates,
for which the coherent dynamics in the tight direction can be neglected.
Here, cylindrical coordinate ${\bm r} \! = \! ( r, \theta, z)$ is introduced and the integration in the two-dimensional plane is denoted as
$\int_{\rm 2D} d{\bm r} \! \equiv \! \int r dr \int \sin \theta d\theta$.
The total number of the atoms
$N \! \equiv \! \sum_i \int_{\rm 2D} d{\bm r} |\phi_i|^2 \! = \! 1.5 \! \times \! 10^3 d^{-1}$
is fixed.
With this set of values of $g_n'$ and $N$, doubly quantized vortex states in scalar BECs have a dynamical instability \cite{pu, mottonen, huhtamaki1, lundh}.
The magnetization is obtained from $M \! \equiv \! \int_{\rm 2D} d{\bm r} \left (|\phi_1|^2 \! - \! |\phi_{-1}|^2 \right ) / N$.

For low enough rotation frequencies, vortex lattices do not form,
and hence Eq.~(\ref{eq:symeq}) holds.
Thus the effect
of the external rotation can be taken into account as a chemical
potential shift such that $\mu_j' \! \equiv \! \mu' \! + \! j
\delta \mu'$, $\mu' \! \equiv \! \mu \! + \hbar \Omega$, and
$\delta \mu' \! \equiv \! \delta \mu \! - \! \hbar \Omega$.
In experiments, the magnetization per particle $M$ is an observable,
and hence the chemical potentials $\mu_i$ can be treated as
Lagrange multipliers in the calculation.
Thus, the rotation
cannot change the Gross-Pitaevskii (GP) solution under constant
$M$. On the other hand, the external rotation changes the
excitation spectrum by $\Delta E_{\bf q} (\Omega) \! \equiv \!
E_{\bf q} (\Omega) \! - \! E_{\bf q} (\Omega \! = \! 0) \! = \! -
\hbar \Omega q_\theta$.


\section{Results}
\label{sec:result}

\subsection{Coreless vortex states}
\label{subsec:coreless} 
\begin{figure}[h]
\includegraphics[width=\linewidth]{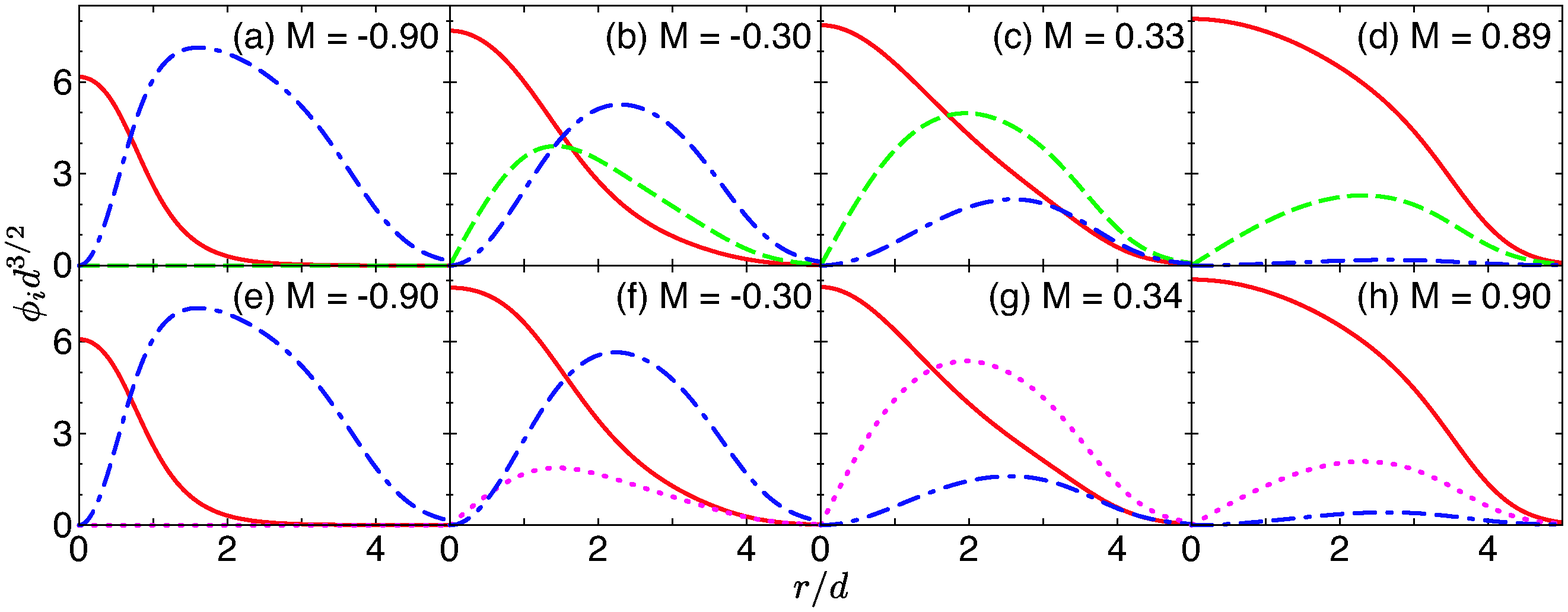}
\caption{(color online) The spatial profile of the order parameter for $g_s' \! = \! -0.001$ (a)--(d) and for $g_s' \! = \! 0.001$ (e)--(h). The magnetization $M$ is (a) $-0.90$, (b) $-0.30$, (c) $0.33$, (d) $0.89$, (e) $-0.90$, (f) $-0.30$, (g) $0.34$, and (h) $0.90$. The solid, dashed, and dashed-dotted lines correspond to $m_F \! = \! 1$, $0$, and $-1$ components, respectively. The order parameter corresponding to $m_F \! = \! 0$ component in the antiferromagnetic case, denoted by the dotted line, is purely imaginary.
\label{fig:wfprofile}
}
\end{figure}

\begin{figure}[h]
\includegraphics[width=0.75\linewidth]{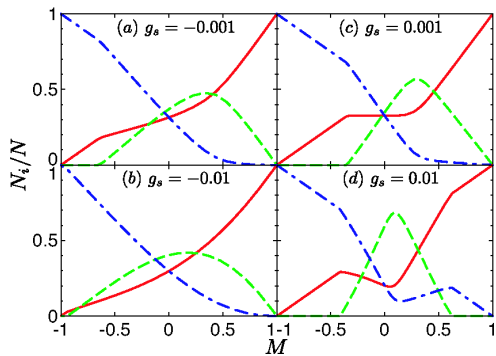}
\caption{(color online)
The ratio of the atoms in hyperfine spin states and total number of the atoms $N_i / N$ as a function of magnetization for $g_s' \! = \! -0.001$ (a), $g_s' \! = \! -0.01$ (b), $g_s' \! = \! 0.001$ (c), and $g_s' \! = \! 0.01$ (d). The solid, dashed, and dashed-dotted lines correspond to $m_F \! = \! 1$, $0$, and $-1$ components, respectively. The total number of atoms in the 2D plane is fixed to $N \! = \! 1.5 \! \times \! 10^3 d^{-1}$.
\label{fig:N_Mdep}
}
\end{figure}

We study the coreless vortex states,
defined by the combination of the phase windings of each component
$\langle w_1, w_0, w_{-1} \rangle \! = \! \langle 0, 1, 2
\rangle$, for magnetization ranging from $-1$ to $1$, and for
different strengths of the spin-spin interaction. In
Fig.~\ref{fig:wfprofile}, we display typical spatial profiles of
the order parameter for $g_s' \! = \! - 0.001$ and $0.001$. In
Fig.~\ref{fig:N_Mdep}, the particle number $N_i$ in different
hyperfine spin states is presented as a function of $M$ for
different values of the spin-spin coupling constant $g_s'$.

According to Isoshima {\it et al.} \cite{isoshima5}, the spin-dependent term of the energy density functional can be written as
\begin{eqnarray}
E_s (r)
&\equiv& \frac{g_s'}{2} \bigl \{
2 \phi_0'^2 (r) [\phi_1' (r) \pm \phi_{-1}' (r)]^2 \nonumber \\
&& \mbox{} + [\phi_1'^2 (r) - \phi_{-1}'^2 (r)]^2 \bigr \}.
\label{eq:pureinteraction}
\end{eqnarray}
Here we have assumed the phase condition $\gamma_1 \gamma_{-1} \gamma_0^{\ast 2} \! = \! \pm 1$ which stems from the requirement that the spin-dependent part of the total energy is minimized. The upper (lower) sign corresponds to ferromagnetic (antiferromagnetic) interaction.
Equation~(\ref{eq:pureinteraction}) helps to understand the $M$
dependence of the order parameter for different values of $g_s'$.
In terms of Eq.~(\ref{eq:pureinteraction}), a large magnitude of
the spin vector is more favorable in the ferromagnetic case, and
oppositely, in the antiferromagnetic case, the spin vector tends
to vanish. By comparing panels (b) and (f) in
Fig.~\ref{fig:wfprofile}, we observe that $\phi_0$ has larger
amplitude in the ferromagnetic case and therefore enhances the magnitude of the
spin vector. Furthermore, for a broad range of $M$, $N_0$ is finite
in the ferromagnetic case whereas it typically vanishes for
antiferromagnetic interactions as shown in Fig.~\ref{fig:N_Mdep}.
Moreover, the fact that the $m_F \! = \! -1$ component has a
different winding number to $m_F \! = \! 1$ component explains the
asymmetry of the distributions in Fig.~\ref{fig:N_Mdep}.

\subsection{Excitation spectra and complex-frequency modes}
\label{subsec:excitation}
We present a typical excitation spectrum to explain the mechanism
behind the appearance of the dynamical instabilities.
As observed from Eq.~(\ref{eq:linearres}), the fluctuation term grows
exponentially in time when some eigenvalue $E_{\bf q}$ is complex. This is
referred to as the dynamical instability.
In such case, small perturbations about the stationary solution of
the GP equation can render it to decay into another state even in
the absence of dissipation.

\begin{figure}[h]
\includegraphics[width=0.8\linewidth]{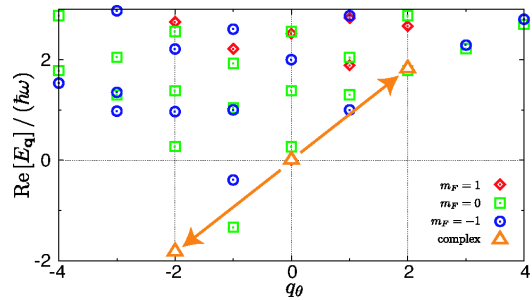}
\caption{(color online) The excitation spectrum for $M \! = \!
-0.9$ and $g_s' \! = \! -0.001$. The triangle at ($q_\theta \! =
\! 0$, ${\rm Re}[E_{\bf q}] \! = \! 0$) corresponds to the GP
solution. The labels indicate the majority component in the
quasiparticle amplitudes $u_i$ and $v_i$. This spectrum includes
the complex-frequency modes denoted by the triangles at ($q_\theta
\! = \! \pm 2$, ${\rm Re}[E_{\bf q}] \! = \! \pm 1.82$).
\label{fig:fullspectrum} }
\end{figure}

A typical excitation spectrum including complex-frequency modes is
shown in Fig.~\ref{fig:fullspectrum}.
In this figure, the horizontal axis is the angular momentum quantum number $q_\theta$
of the excited state and the vertical axis is the real part of the excitation energy
${\rm Re}[E_{\bf q}]$. The eigenstate at $q_\theta \! = \! 0$ and
$E_{\bf q} \! = \! 0$ corresponds to the GP solution.
The
excitation modes are labeled by the majority component of the
excitation, that is, for the excitation with label $m_F \! = \! i$, 
the largest amplitude of the fluctuation is given by $\int_{\rm 2D}
d{\bm r} \left [ | u_i |^2 \! + \! | v_i |^2 \right ]$.

The excitation spectrum shown in Fig.~\ref{fig:fullspectrum} corresponds to $g_s' \! = \! -0.001$ and $M \! = \! -0.9$. In this case, the state derived from the GP equation has most of the particles occupying the $m_F \! = \! -1$ component with a winding number $w_{-1} \! = \! 2$ and a small amount of $m_F \! = \! 1$ component fills the core of the vortex in the $\phi_{-1}$ component, see Fig.~\ref{fig:wfprofile}(a).

The labels also illustrate the nature of the excitation modes.
For example, in the $M \! = \! -1$ limit, the $m_F \! = \! 1$, $0$, and $-1$ modes correspond to longitudinal spin fluctuations, transverse spin fluctuations, and density fluctuations.
However, apart from this limit, the excitation modes are more complicated, because of the mixing between different spin components.

In the spontaneous dynamical excitation of the complex-frequency
modes, conservation of the total energy and angular momentum must
be satisfied.
As depicted in Fig.~\ref{fig:fullspectrum}, the pair of
complex-frequency modes with ($q_\theta \! = \! -2$, ${\rm Re}
[E_{\bf q}] \! = \! -1.82$) and ($q_\theta \! = \! 2$, ${\rm Re}
[E_{\bf q}] \! = \! 1.82$) satisfies the aforementioned constraints,
and hence the initial state with ($q_\theta \! = \! 0$, ${\rm
Re}[E_{\bf q}] = 0$) can spontaneously decay into these two states
without any dissipation. There are also additional restrictions
for the appearance of the complex-frequency modes which will be
discussed later.
We also note that external rotation does not affect this condition
since the excitation energies are shifted by $-\hbar \Omega
q_\theta$, see Sec.~\ref{sec:formulation}.

\begin{figure}[h]
\includegraphics[width=0.85\linewidth]{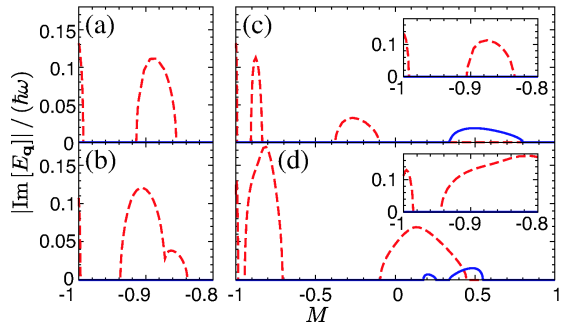}
\caption{(color online)
The absolute values of the imaginary parts
of the complex-frequency eigenvalues are shown as a function of the
magnetization $M$ for $g_s' \! = \! -0.001$ (a), $g_s' \! = \!
-0.01$ (b), $g_s' \! = \! 0.001$ (c), and $g_s' \! = \! 0.01$ (d). The
$q_\theta \! = \! \pm 2$ and $q_\theta \! = \! \pm 1$ modes are
indicated by dashed and solid lines, respectively. The insets in
(c) and (d) show the detailed structure in the vicinity of $M \! =
\! -1$ for $g_s' \! = \! 0.001$ and $g_s' \! = \! 0.01$. There are
no complex-frequency modes found for $M \! > \! -0.8$ in the
ferromagnetic case.
\label{fig:imaginarypart} }
\end{figure}

Several complex-frequency modes are found in both ferromagnetic
and antiferromagnetic cases.
Figure \ref{fig:imaginarypart} presents the imaginary part of the
eigenvalues as a function of $M$. We find that two types of
complex-frequency modes can appear in the coreless vortex states:
(i) a pair of $q_\theta \! = \! \pm 2$ modes,
and (ii) a pair of $q_\theta \! = \! \pm 1$ modes. The former
complex-frequency mode appears in the vicinity of $M \! = \! -1$
in both ferromagnetic and antiferromagnetic cases as shown in
Fig.~\ref{fig:imaginarypart}. The results in the $M \! = \! -1$
limit reproduce those of the doubly quantized vortex in a scalar
BEC \cite{pu, mottonen, huhtamaki1, lundh, kawaguchi}. In contrast, another
pair of complex-frequency modes with $q_\theta \! = \! \pm 1$
emerges in the antiferromagnetic interaction regime.

\begin{figure}[h]
\includegraphics[width=0.95\linewidth]
{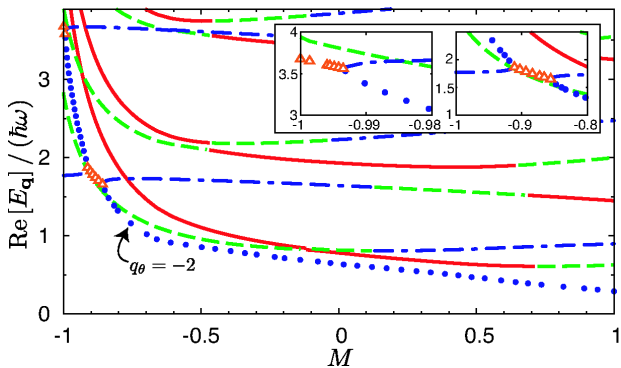}
\caption{(color online) The magnetization dependence of the excitation energies of
the $q_\theta \! = \! \pm 2$ modes for $g_s' \! = \! - 0.001$.
The insets show the details of the spectrum where complex eigenenergies appear.
The eigenenergy of the $q_\theta \! = \! -2$ mode is plotted as $- {\rm Re}[E_{\bf q}]$.
The majority component in the quasiparticle amplitudes $u_i$ and $v_i$ of the corresponding excitation energies is indicated by the solid,
dashed, and dashed-dotted lines for the $m_F \! = \! 1$, $0$, and $-1$
components in $q_\theta \! = \! 2$ mode, respectively. The dots
indicate $q_\theta \! = \! -2$ mode, which is dominated by the
$m_F \! = \! -1$ component. The complex-frequency modes are
labeled by triangles in both $q_\theta \! = \! \pm 2$ modes. }
\label{fig:mdep_qt-2gs-0.001}
\end{figure}

Figure~\ref{fig:mdep_qt-2gs-0.001} shows the excitation energies
of the $q_\theta \! = \! \pm 2$ modes for $g_s' \! = \! -0.001$ as
a function of $M$. The solid, dashed, and dashed-dotted lines
correspond to $q_\theta \! = \! 2$ modes, for which the majority
components are $m_F \! = \! 1$, $0$, and $-1$, respectively. The
eigenenergy of the $q_\theta \! = \! -2$ mode is plotted as $-
{\rm Re} [ E_{\bf q} ]$ and denoted by dots. The majority
component for this excitation is $m_F \! = \! -1$.
The complex-frequency modes appear in the regions where $q_\theta
\! = \! 2$ and $q_\theta \! = \! -2$ modes overlap,
due to the energy and angular momentum constraints.

Let us discuss the dependence of $q_\theta \! = \! 2$ modes shown in
Fig.~\ref{fig:mdep_qt-2gs-0.001}.
These modes are classified as quadrupole modes, which give rise to
the two-fold rotational symmetric deformation of the condensate.
In spinor BECs, due to the multicomponent sublevels of the order parameter,
there are three kinds of quadrupole modes: the transverse and longitudinal
spin quadrupole modes and the density quadrupole mode,
which correspond to
the three lowest lines around $M \! = \! 1$ in Fig.~\ref{fig:mdep_qt-2gs-0.001},
respectively.
The other modes with higher energy are the higher order quadrupole modes.
The lowest density fluctuation mode with $q_\theta \! = \! 2$ is embed at
$E_{\bf q} \! = \! 1.45 \hbar \omega$ around $M \! = \! 1$, which is in good agreement with
$E_{\bf q} \! = \! \sqrt{2} \hbar \omega$ derived within the Thomas-Fermi approximation
\cite{stringari}.
With increasing $M$, since the ground state has a finite angular momentum $\langle l_z \rangle$
associated with the windings $\langle 0, 1, 2 \rangle$, the energy gradually shifts as
$E_{\bf q} (M) \! - \! E_{\bf q} (M \! = \! 1) \! \propto \! \langle l_z \rangle$ \cite{zambelli}
and stays around $E_{\bf q} \! = \! 1.5 \hbar \omega$ in the whole $M$ region.
We also note that the energy of the transverse and longitudinal quadrupole modes,
which are shown with solid and dashed lines near $M \! = \! -1$,
rapidly increase as $M$ decreases
because of the increase of the relative chemical potential difference $\delta \mu$.
Since the energy of the lowest excitations with
$q_\theta \! = \! -2$ increases with $M$ near $M \! = \! -1$ and
the resonating $q_\theta \! = \! 2$ density quadrupole mode
remains almost constant,
the complex-frequency modes eventually disappear, as shown in
Fig.~\ref{fig:mdep_qt-2gs-0.001}. The complex-frequency modes
appear again near $M \! = \! - 0.9$ since the $q_\theta \! = \!
-2$ excitation mode finds another mode to pair with such that the
total energy and angular momentum conservations are satisfied.

\begin{figure}[h]
\includegraphics[width=0.9\linewidth]
{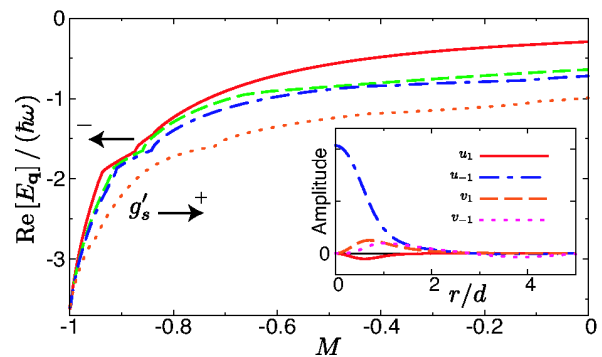}
\caption{(color online) The magnetization dependence of the lowest energy excitations with $q_\theta \! = \! -2$ for different values of $g_s'$. The solid, dashed, dashed-dotted, and dotted lines indicate $g_s' \! = \! -0.01$, $-0.001$, $0.001$, and $0.01$ cases, respectively.
The shift towards positive energy increases with $g_s'$ decreasing from positive values to negative values.
The inset shows the absolute values of the quasiparticle amplitudes $\{ u_i, v_i \}$
for $g_s \! = \! -0.001$ and $M \! = \! -0.95$.
The $u_0$ and $v_0$ are neglected since they are vanishingly small.
}
\label{fig:lowest}
\end{figure}

The complex-frequency modes appear for a clearly wider range of
values of $M$ in the antiferromagnetic case compared with the
ferromagnetic case [see Fig.~\ref{fig:imaginarypart}(a), (b) and
insets of panels (c) and (d)]. To explain this tendency, we
consider the lowest negative energy excitation with $q_\theta \! = \! -2$.
The excitation energy increases faster with increasing $M$
in the ferromagnetic case as shown in Fig.~\ref{fig:lowest}.
This tendency results from the spatial
profile for the $q_\theta \! = \! -2$ mode. The lowest energy
excitation with $q_\theta \! = \! -2$ is mainly composed of the
$u_{-1}$ wave function as shown in the inset of
Fig.~\ref{fig:lowest}. The excitation wave function can be
generally expanded in terms of the $q$th Bessel function $J_q (r)$ as
$u_{{\bf q}, i} (r) \! = \! \sum_{s \! = \! 1}^{\infty} A_s
J_{q_\theta \! + \! w_i} (k_s r)$, where $k_s \! = \! \lambda_s /
L$. Here $\lambda_s$ is the zero point of the Bessel function and
$L$ is the cutoff length of the system. Since the Bessel function
behaves as $J_q (r) \! \sim \! r^{|q|}$ near $r \! = \! 0$, and
$q_\theta \! + \! w_{-1} \! = \! 0$,
we have $u_{-1} \! \propto \! r^0$ near $r \! = \! 0$.
Hence, the lowest eigenmode at $q_\theta \! = \! -2$ is the core localized mode and the quasiparticle amplitude $u_{-1} (r)$ spatially overlaps with $\phi_1$, and fills the vortex core.
Due to the coupling term $-g_s |\phi_1|^2 u_{-1}$ in the BdG matrix~(\ref{eq:bdgmatrix}),
the lowest eigenvalue increases rapidly in the ferromagnetic regime.
In addition, the slope near $M \! = \! -1$ in
Fig.~\ref{fig:lowest} is steeper in the ferromagnetic case than in
the antiferromagnetic case. The $q_\theta \! = \! 2$ modes with
positive ${\rm Re} [ E_{\bf q} ]$ in resonance with lowest $q_\theta
\! = \! -2$ mode are less sensitive to changes in $M$ as we have discussed above.
Hence, the
complex-frequency eigenmode can appear only in narrow
magnetization regions in the case of ferromagnetic interactions. Apart
from $M \! \sim \! -1$, in the ferromagnetic case, the coreless
vortex becomes dynamically stable.

\begin{figure}[h]
\includegraphics[width=0.7\linewidth]{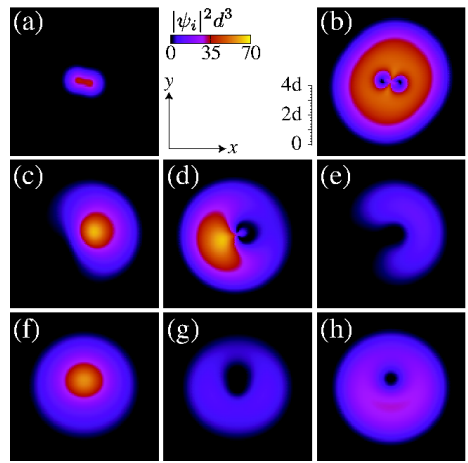}
\caption{(color online)
The density profiles $|\Psi_1|^2$ (left column),
$|\Psi_0|^2$ (center column), and $|\Psi_{-1}|^2$ (right column)
perturbed by excitation modes.
The upper row corresponds to the complex-frequency $q_\theta \! = \! \pm 2$ modes
for $g_s' \! = \! -0.001$ and $M \! = \! -0.9$,
the middle row to of complex-frequency $q_\theta \! = \! \pm 1$ modes
for $g_s' \! = \! 0.01$ and $M \! = \! 0.2$, and the lower row
to the lowest real-frequency $q_\theta \! = \! -1$ mode for $g_s' \! = \! 0.01$ and $M \! = \! 0.3$.
The density profile of the $m_F \! = \! 0$ component $|\Psi_0|^2$
in upper row is neglected since it is vanishingly small.
The complex modes with both positive and negative $q_\theta$ are equally superposed
since the modes appear as a result of the energy and angular momentum conservation.
The field of view is $10 d \! \times \! 10 d$.
We take $\lambda \! = \! 100$ to show the essential qualitative features of the 
fluctuation.
\label{fig:densityfluctuation}
}
 \end{figure}

Let us consider the difference in the density fluctuations induced
by the two complex-frequency modes. The perturbed density profile
of each component is shown in Fig.~\ref{fig:densityfluctuation},
where the first and second rows show the density fluctuations
caused by the complex-frequency modes with $q_\theta \! = \! \pm 2$
and $q_\theta \! = \! \pm 1$, respectively. From the left to the
right column, the density of the $m_F \! = \! 1, 0, -1$ components
are shown. The $m_F \! = \! 0$ component of the $q_\theta \! = \!
\pm 2$ complex-frequency mode is neglected, because its amplitude is
vanishingly small.

The $q_\theta \! = \! \pm 2$ complex-frequency mode breaks the doubly
quantized vortex in the $m_F \! = \! -1$ component into two singly
quantized vortices, as shown in Fig.~\ref{fig:densityfluctuation}(b). This
mechanism of dynamical instability is equivalent to the dynamical
instability of a doubly quantized vortex in scalar BECs. It has
also been found in the studies of coreless vortices induced by
external magnetic fields \cite{ville}. On the other hand, the
$q_\theta \! = \! \pm 1$ complex-frequency mode,
which appears only in the antiferromagnetic regime, has a fundamentally
different response on the condensate. We categorize this kind of
dynamical instability mode as phase separation, since this mode
leads to a spatial separation of the could into domains of a
certain component $\phi_{1}$, $\phi_{-1}$ or $\phi_0$. Although
the separation is not very sharp, it is clearly visible in
Fig.~\ref{fig:densityfluctuation}. Furthermore, the $m_F \! = \!
1$ and $m_F \! = \! -1$ components tend to spatially overlap with
each other, which is attributed to the attractive interaction
between them due to the antiferromagnetic interaction,
as seen in Eq.~(\ref{eq:pureinteraction}).

\subsection{Stable modes}
\label{subsec:stable}
\begin{figure}[h]
\includegraphics[width=0.9\linewidth]{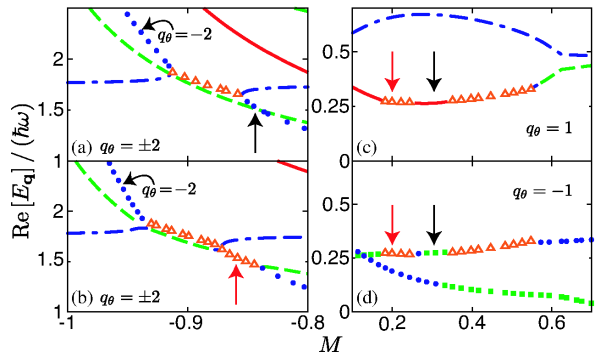}
\caption{(color online) The detailed structures of the excitation spectra. We show $q_\theta \! = \! \pm 2$ modes for $g_s' \! = \! -0.001$ (a) , and $g_s' \! = \! -0.01$ (b), and
 $q_\theta \! = \! 1$ (c), and $q_\theta \! = \! -1$ (d) modes for $g_s' \! = \! 0.01$. The eigenenergies for $q_\theta \! = \! -2$ and $q_\theta \! = \! -1$ are plotted as $- {\rm Re}[E_{\bf q}]$. For $q_\theta \! = \! 2$ and $q_\theta \! = \! 1$ modes, the majority components of the spectrum are indicated by solid, dashed, and dashed-dotted lines corresponding to $m_F \! = \! 1$, $0$, and $-1$, respectively. For the $q_\theta \! = \! -2$ and $q_\theta \! = \! -1$ modes, these are indicated by filled squares and dots corresponding to $m_F \! = \! 0$, and $-1$ components, respectively. The complex-frequency modes are labeled by triangles in all cases.
\label{fig:pick}
}
\end{figure}

In addition to the dynamical instabilities, there are modes with real 
eigenvalues even if the restrictions of the conservation of the total energy and angular momentum are satisfied.
For example, (i) $q_\theta \! = \! \pm 2$ modes for $g_s' \! = \! -0.001$ near $M \! = \! -0.85$
are shown in Fig.~\ref{fig:pick}(a), and (ii) $q_\theta \! = \! \pm 1$ modes for
$g_s' \! = \! 0.01$ near $M \! = \! 0.3$ in Fig.~\ref{fig:pick}(c) and (d).

Let us first consider the case (i). The corresponding modes have $m_F
\! = \! -1$ component in majority for the $q_\theta \! = \! -2$
mode and $m_F \! = \! 0$ for the $q_\theta \! = \! 2$ mode. In
particular, they satisfy the condition of the conservation of the
total energy and angular momentum.
Thus they can in principle form an excitation with complex
eigenvalue, and in fact, this is the case for $g_s' \! = \! -0.01$
and certain values of $M$ as shown in Fig.~\ref{fig:pick}(b). The
difference between these two cases can be traced back to the
stationary solution of the GP equation.
According to Fig.~\ref{fig:N_Mdep}(a) and (b), $\phi_0$ remains
negligible for $M \! \lesssim \! -0.65$ and $g_s' \! = \! -0.001$,
but $\phi_0$ is finite in the overlapping region for $g_s' \! = \! -0.01$.
Hence the existence of a finite number atoms in the corresponding
$q_\theta \! > \! 0$ mode can be considered as another restriction for the
appearance of the dynamical instability.

Next, we move to the case (ii). In Fig.~\ref{fig:pick}(c) and (d),
we show excitations with purely real eigenfrequencies at $M \!
\sim \! 0.3$ surrounded by excitations corresponding to complex
eigenvalues. To understand this behavior, we consider the related
density fluctuations. The density profiles in the case (ii) are
shown in Fig.~\ref{fig:densityfluctuation}(f)--(h). We notice that
the fluctuation leads to the precession motion of the
$\left \langle 0, 1, 2 \right \rangle$ coreless vortex and
the phase separation appearing for $M \! = \! 0.2$
[Fig.~\ref{fig:densityfluctuation}(c)--(e)] does not occur here.
Hence we argue that the phase separation is a characteristic
feature of this particular type of dynamical instability.

In addition to the above discussion on the existence of
complex-frequency modes, we note that the existence of dipole
modes is a general feature of the excitation spectrum of a
harmonically trapped many-particle system. A generalization
\cite{dobson,fetter} of the Kohn's theorem \cite{kohn} shows that
these center-of-mass oscillation modes should exist for scalar
particles with energy eigenvalue $E_{\bf q} \! = \! \hbar \omega$
independent of the interaction strength. Thus the existence of
Kohn modes in the theory describing the system is typically used
to check for the validity of the approximations made. It turns out
that the dipole modes have exactly the energy $\hbar\omega$ in the
finite-temperature Bogoliubov approximation, in which the spatial
dependence of thermal gas component is neglected in the GP and BdG
equations. For so-called Popov and second-order finite-temperature
mean-field theories, the excitation energy is very close to,
although not exactly, the trap energy~\cite{mottonen3}. In
Appendix A, we present a proof that Kohn modes with energy $E_{\bf
q} \! = \! \hbar \omega$ exist for the BdG equations we utilize
independent of the magnetization, density-density, or spin-spin
interactions.

\section{Conclusions}
\label{sec:conclusion}
We have investigated the stability of the
coreless vortex states in $F \! = \! 1$ spinor Bose-Einstein
condensates. Namely, we have calculated the low-energy excitation
spectra in the whole range of magnetization $M$ by solving the
Gross-Pitaevskii and the Bogoliubov-de Gennes equations. The
complex-frequency modes, which cause the dynamical instabilities, have
been found in both ferromagnetic and antiferromagnetic cases.

The complex-frequency modes in the ferromagnetic case cause the
doubly quantized vortex to decay into a pair of singular vortices.
In addition, antiferromagnetic interactions were found to cause
phase separation through dynamical instability of coreless
vortices.
In general, we found that the dynamical instabilities tend to be
suppressed by the ferromagnetic interactions and oppositely
enhanced by the antiferromagnetic ones. We also note that rather
slow external rotation does not have an effect on the dynamical
instabilities for a fixed magnetization.

In addition to the conventional energy and angular momentum
conservation, we found other restrictions for the appearance of
the dynamical instability. One such a restriction for the
$q_\theta \!
> \! 0$ mode is the need for a considerable particle number
in the component of the condensate order parameter to be excited.
Furthermore, we found that only certain $q_\theta \! < \! 0$
modes can resonate with other modes. These correspond to the vortex splitting
mode with $q_\theta \! = \! -2$ in both interaction regimes,
and the phase separating mode with $q_\theta \! = \! -1$ in the
antiferromagnetic regime.
Due to these constraints, a dynamically stable coreless vortex can
exist for certain magnetizations $M$, not only in the
ferromagnetic case but also in the antiferromagnetic case.
Our studies can be verified experimentally in fully optically
trapped spinor BECs using present-day techniques.

\section*{ACKNOWLEDGMENTS}

We thank M.~Mine and J.~A.~M.~Huhtam{\" a}ki for useful discussions.
This work was supported by
a grant of the Japan Society for the Promotion of Science (MT, TM, and KM),
the Jenny and Antti Wihuri Foundation (VP),
the Academy of Finland (MM),
and Emil Aaltonen's Foundation (MM and VP).

\appendix
\section{Existence of Kohn modes}
Here, we show that dipole modes exist in harmonically trapped
spinor Bose-Einstein condensates described by the employed
mean-field theory. We consider a general system at zero
temperature.
The single particle Hamiltonian is defined as,
\begin{eqnarray}
H_0 ({\bm r})
&=& - \frac{\hbar^2}{2 m} \nabla^2 + V_{\rm trap} ({\bm r}),
\end{eqnarray}
where $V_{\rm trap} ({\bm r})$ is a general three-dimensional
harmonic trap potential,
$V_{\rm trap} ({\bm r}) \! = \! \frac{1}{2} m \sum_\alpha
\omega_\alpha^2 \alpha^2$,
where $\alpha$ takes values $x$, $y$, and $z$. We introduce the
following creation and annihilation operators,
\begin{eqnarray}
\left . \begin{array}{c} a_\alpha^\dagger \equiv \frac{1}{\sqrt 2}
\left ( \frac{\alpha}{d_\alpha}
- d_\alpha \frac{\partial}{\partial \alpha} \right ), \\
a_\alpha
\equiv \frac{1}{\sqrt 2} \left ( \frac{\alpha}{d_\alpha}
+ d_\alpha \frac{\partial}{\partial \alpha} \right ),
\end{array} \right .
\end{eqnarray}
where $d_\alpha \! \equiv \! \sqrt{\hbar / m \omega_\alpha}$.
The introduced operators satisfy the bosonic commutation relation,
\begin{eqnarray}
[ a_\alpha, a_{\alpha'}^\dagger ]
= \delta_{\alpha, \alpha'},
\hspace{5mm}
[ a_\alpha, a_\alpha ]
= [ a_\alpha^\dagger, a_{\alpha'}^\dagger]
= 0.
\end{eqnarray}
Using this notation, the single particle Hamiltonian can be written in the form,
\begin{eqnarray}
H_0 ({\bm r})
&=& \sum_\alpha \hbar \omega_\alpha \left (
a_\alpha^\dagger a_\alpha + \frac{1}{2} \right ).
\end{eqnarray}

We denote the order parameter for an arbitrary spin $F$ BEC
with the $(2 F \! + \! 1)$-dimensional vector,
\begin{eqnarray}
{\bf \Psi} ({\bm r})
&=& \left [ \Psi_F ({\bm r}), \Psi_{F - 1} ({\bm r}),
\cdots, \Psi_{-F} ({\bm r}) \right ]^T.
\label{eq:generalorderparameter}
\end{eqnarray}
The GP equation can be written in a general form,
\begin{eqnarray}
\left [ H_0 ({\bm r}) \underline{\tau_0}
+ {\underline \Sigma} ({\bm r}) \right ]
{\bf \Psi} ({\bm r}) + \tilde{\underline{\Delta}} ({\bm r})
{\bf \Psi}^\ast ({\bm r})
&=& \mu {\bf \Psi} ({\bm r}).
\label{eq:gpsimple}
\end{eqnarray}
Here the $(2 F \! + \! 1)$-dimensional square matrices
${\underline \Sigma} ({\bm r})$ and $\tilde{\underline{\Delta}}
({\bm r})$ are local selfenergies. A $(2 F \! + \! 1)$-dimensional
unit matrix $\underline{\tau_0} \! \equiv \! {\rm diag} ( 1,
\cdots, 1)$ is also introduced.
From Eq.~(\ref{eq:gpsimple}) we obtain a set of two equations
\begin{eqnarray}
\lefteqn{
\left [ \hat{H_0} ({\bm r}) - \hbar \omega_\alpha \hat{\tau_0} \right ]
\left [ \begin{array}{c}
a_\alpha^\dagger {\bf \Psi} ({\bm r}) \\
a_\alpha {\bf \Psi}^\ast ({\bm r})
\end{array} \right ]} \nonumber \\
&=& \left [ \begin{array}{c}
- \{ a_\alpha^\dagger \underline{\Sigma} ({\bm r}) \} {\bf \Psi} ({\bm r}) \\
\{ a_\alpha \underline{\Sigma}^\ast ({\bm r}) \} {\bf \Psi}^\ast ({\bm r})
\end{array} \right ]
+ \left [ \begin{array}{c}
- a_\alpha^\dagger \{ \tilde{\underline{\Delta}} ({\bm r}) \}
{\bf \Psi}^\ast ({\bm r}) \\
a_\alpha \{ \tilde{\underline{\Delta}}^\ast ({\bm r}) {\bf \Psi} ({\bm r}) \}
\end{array} \right ].
\label{eq:GP}
\end{eqnarray}
Here we introduce $\hat{\tau_0} \! \equiv \! {\rm diag} (\underline{\tau_0}, \underline{\tau_0})$ and a $2 \! \times \! (2 F \! + \! 1)$-dimensional square matrix $\hat{H_0}$
\begin{eqnarray}
\hat{H_0} ({\bm r})
&\equiv& {\rm diag} \Bigl [ \{ H_0 ({\bm r}) - \mu \} \underline{\tau_0}
+ \underline{\Sigma} ({\bm r}), \nonumber \\
&& - \{ H_0 ({\bm r}) - \mu \} \underline{\tau_0}
- \underline{\Sigma}^\ast ({\bm r}) \Bigr ].
\end{eqnarray}

From Eq.~(\ref{eq:generalorderparameter}) one can derive the general form of the BdG equation
\begin{eqnarray}
\hat{H_0} \left [ \begin{array}{c}
{\bm u}_\nu ({\bm r})\\
{\bm v}_\nu ({\bm r})
\end{array} \right ]
+ \left [ \begin{array}{c}
\underline{\Delta} ({\bm r}) {\bm v}_\nu ({\bm r}) \\
- \underline{\Delta}^\ast ({\bm r}) {\bm u}_\nu ({\bm r})
\end{array} \right ]
= E_\nu \left [ \begin{array}{c}
{\bm u}_\nu ({\bm r}) \\
{\bm v}_\nu ({\bm r})
\end{array} \right ].
\end{eqnarray}
At zero temperature, $\underline{\Delta} ({\bm r})$ is equal to
$\underline{\tilde{\Delta}} ({\bm r})$ in the GP equation. Let us
take an ansatz
\begin{eqnarray}
\left [ \begin{array}{c}
{\bm u}_{\nu = \alpha} ({\bm r}) \\
{\bm v}_{\nu = \alpha} ({\bm r})
\end{array} \right ]
&=& \left [ \begin{array}{c}
a_\alpha^\dagger {\bf \Psi} ({\bm r}) \\
a_\alpha {\bf \Psi}^\ast ({\bm r})
\end{array} \right ],
\end{eqnarray}
and write the BdG equation using Eq.~(\ref{eq:GP}) in the form
\begin{widetext}
\begin{eqnarray}
\left [ E_\alpha - \hbar \omega_\alpha \right ]
\left [ \begin{array}{c}
a_\alpha^\dagger {\bf \Psi} ({\bm r}) \\
a_\alpha {\bf \Psi}^\ast ({\bm r})
\end{array} \right ]
&=& \left [ \begin{array}{c}
- \{ a_\alpha^\dagger \underline{\Sigma} ({\bm r}) \} {\bf \Psi} ({\bm r})
- a_\alpha^\dagger \{ \underline{\Delta} ({\bm r}) {\bf \Psi}^\ast ({\bm r}) \}
+ \underline{\Delta} ({\bm r}) a_\alpha {\bf \Psi}^\ast ({\bm r}) \\
\{ a_\alpha \underline{\Sigma}^\ast ({\bm r}) \} {\bf \Psi}^\ast ({\bm r})
+ a_\alpha \{ \underline{\Delta}^\ast ({\bm r}) {\bf \Psi} ({\bm r}) \}
- \underline{\Delta}^\ast ({\bm r}) a_\alpha^\dagger {\bf \Psi} ({\bm r})
\end{array} \right ] .
\label{eq:BdG}
\end{eqnarray}
\end{widetext}

All results above are for a general BEC with hyperfine spin $F$.
Below, we restrict the discussion to $F \! = \! 1$ case since the selfenergy for this case is known.
Here, the order parameter takes the form
${\bf \Psi} ({\bm r}) \! = \! [ \Psi_1 ({\bm r}), \Psi_0 ({\bm r}), \Psi_{-1} ({\bm r}) ]^T$.
Using the following notation \cite{kondo},
\begin{eqnarray}
\underline{A}^\nu
&\equiv& \left \{ \begin{array}{lll}
\underline{\tau_0} & {\rm for} & \nu = 0 \\
\underline{F}^\nu & {\rm for} & \nu = 1, 2, 3
\end{array} \right .,
\end{eqnarray}
\begin{eqnarray}
g_\nu
&\equiv& \left \{ \begin{array}{lll}
g_n' & {\rm for} & \nu = 0 \\
g_s' & {\rm for} & \nu = 1, 2, 3
\end{array} \right .,
\end{eqnarray}
the selfenergies are written as,
\begin{eqnarray}
\underline{\Sigma} ({\bm r})
&=& g_\nu \left [ {\bf \Psi}^{\dagger} ({\bm r}) \underline{A}^\nu
{\bf \Psi} ({\bm r}) \underline{A}^\nu
\! + \! \underline{A}^\nu {\bf \Psi} ({\bm r})
{\bf \Psi}^\dagger \underline{A}^\nu \right ],
\end{eqnarray}
\begin{eqnarray}
\underline{\Delta} ({\bm r})
&=& - g_\nu \underline{A}^\nu {\bf \Psi} ({\bm r}) \left [
{\bf \Psi}^\dagger ({\bm r}) \underline{A}^\nu \right ]^\ast,
\end{eqnarray}
where summation over repeated superscripts is implied.
We substitute these selfenergies to the BdG equation (\ref{eq:BdG}), and using the condition
$\left [ {\bf \Psi}^\dagger \underline{A}^\nu {\bf \Psi} \right
]^\ast \! = \! {\bf \Psi}^\dagger \underline{A}^\nu {\bf \Psi}$ we
finally observe that
\begin{eqnarray}
\left [ E_\alpha - \hbar \omega_\alpha \right ]
\left [ \begin{array}{c}
a_\alpha^\dagger {\bf \Psi} ({\bm r}) \\
a_\alpha {\bf \Psi}^\ast ({\bm r})
\end{array} \right ]
&=& \left [ \begin{array}{c}
{\bm \eta} ({\bm r}) \\
- {\bm \eta}^\ast ({\bm r})
\end{array} \right ],
\label{eq:bdgfinal}
\end{eqnarray}
where we have defined,
\begin{eqnarray}
{\bm \eta} ({\bm r})
&\equiv&
- \{ a_\alpha^\dagger \underline{\Sigma} ({\bm r}) \} {\bf \Psi} ({\bm r})
\! - \! a_\alpha^\dagger
\{ \underline{\Delta} ({\bm r}) {\bf \Psi}^\ast ({\bm r}) \} \nonumber \\
&& \mbox{} + \underline{\Delta} ({\bm r}) a_\alpha {\bf \Psi}^\ast ({\bm r}).
\nonumber
\end{eqnarray}
Assuming that $E_\alpha$ is real, Eq.~(\ref{eq:bdgfinal}) yields
\begin{eqnarray}
( E_\alpha - \hbar \omega_\alpha ) ( a_\alpha ^\dagger + a_\alpha)
{\bf \Psi} ({\bm r}) = {\bf 0}.
\end{eqnarray}
Since $(a_\alpha^\dagger \! + \! a_\alpha) {\bf \Psi} ({\bm r}) \! \ne \! 0$,
we conclude that there always exists a mode with energy $E_\alpha \! = \! \hbar \omega_\alpha$.
Therefore the Kohn mode exists irrespective of the atom-atom interactions.
The eigenvector and eigenenergy are given by
$E_\alpha \! = \! \hbar \omega_\alpha$ and
$[ {\bm u}_\nu, {\bm v}_\nu]^T \! = \! [ a_\alpha^\dagger {\bf \Psi} ({\bm r}), a_\alpha {\bf \Psi}^\ast ({\bm r}) ]^T$, respectively.
In our numerical calculations, we typically find the dipole mode with a relative error is
less than $1.5 \! \times \! 10^{-5}$.


\end{document}